# Project Lyra: Catching 1I/'Oumuamua – Mission Opportunities After 2024

Adam Hibberd[1], Andreas M. Hein[1], T. Marshall Eubanks[1,2]

[1] Initiative for Interstellar Studies (i4is)

Bone Mill, New Street, Charfield, GL12 8ES, United Kingdom

[2] Space Initiatives Inc., USA

## Abstract

In October 2017, the first interstellar object within our solar system was discovered. Today designated 1I/'Oumuamua, it shows characteristics that have never before been observed in a celestial body. Due to these characteristics, an in-situ investigation of 1I would be of extraordinary scientific value. Previous studies have demonstrated that a mission to 1I/'Oumuamua is feasible using current and near-term technologies however with an anticipated launch date of 2020-2021, this is too soon to be realistic. This paper aims at addressing the question of the feasibility of a mission to 1I/'Oumuamua in 2024 and beyond. Using the OITS trajectory simulation tool, various scenarios are analyzed, including a powered Jupiter flyby and Solar Oberth maneuver, a Jupiter powered flyby, and more complex flyby schemes including a Mars and Venus flyby. With a powered Jupiter flyby and Solar Oberth maneuver, we identify a trajectory to 1I/'Oumuamua with a launch date in 2033, a total velocity increment of 18.2 km/s, and arrival at 1I/'Oumuamua in 2048. With an additional deep space maneuver before the powered Jupiter flyby, a trajectory with a launch date in 2030, a total velocity increment of 15.3 km/s, and an arrival at 1I/'Oumuamua in 2052 were identified. Both launch dates would provide over a decade for spacecraft development, in contrast to the previously identified 2020-2021 launch dates. Furthermore, the distance from the Sun at the Oberth burn is at 5 Solar radii. This results in heat flux values, which are of the same order of magnitude as for the Parker Solar Probe. We conclude that a mission to 1I/'Oumuamua is feasible, using existing and near-term technologies and there is sufficient time for developing such a mission.

## 1. Introduction

1I/'Oumuamua is the first interstellar object to be observed within our Solar System [1–3]. Since its discovery in October 2017, 1I/'Oumuamua has generated considerable interest from both academia and the media. The academic debate is still ongoing and topics are as wide-ranging as the object's shape [4–9], its composition [4,5,10–14], its origin [15–21], explanation for an observed acceleration [22,23], and estimates for the population of similar objects [2,5]. Given that 1I/'Oumuamua is the first confirmed interstellar object that has been discovered intercepting our Solar System, it might turn out to be the only opportunity to study interstellar material in-situ for decades or even centuries to come [2,5]. Could a spacecraft be sent to 1I and collect scientific data in-situ? Seligman and Laughlin [24] have previously concluded that missions to objects similar to 1I would be feasible but dependent on an early detection and early launch. Hein et al. [25,26] have argued that a mission to 1I would be feasible, using existing and near-term technologies and a combination of a powered Jupiter flyby and a Solar Oberth maneuver. However, the optimal determined launch date between 2020 and 2021 is deemed to be very challenging from an engineering point of view, as the development of interplanetary spacecraft takes at least 5 and often more than 10 years. Therefore, one of the key questions regarding the feasibility of a mission to 1I is whether or not missions with a launch date at least 5 years into the future (2024 onwards) are feasible in terms of mission duration and velocity requirement $\Delta V$. In this paper we demonstrate that a mission to 1I/'Oumuamua is feasible at later points in time, providing sufficient time for developing a spacecraft.

## 2. Materials and Methods

For finding trajectories with a sufficiently low ΔV and acceptable mission duration, various optimal flyby configurations are analyzed. As a cut-off criteria for ΔV, we select 18.3 km/s from Hein et al. [26]. This ΔV allows for reasonably sized spacecraft masses for existing or near-term launch systems such as the Falcon Heavy and the Space Launch System. For the maximum flight duration from launch to 1I/'Oumuamua encounter, we somewhat arbitrarily select 30 years, as it stays within the career of scientists who have worked on the formulation of the mission at the beginning of their career.

The trajectories are calculated using the Optimum Interplanetary Trajectory Software (OITS) developed by Adam Hibberd, based on a patched conic approximation. With the patched conic approximation, only the gravitational attraction of a celestial body is taken into consideration within its sphere of influence and the gravitational attraction of other bodies is neglected. The trajectory connecting each pair of celestial bodies is determined by solving the Lambert problem using the Universal Variable Formulation [27]. The resulting non-linear global optimization problem with inequality constraints is solved applying the NOMAD solver [28].

In the following investigations, total ΔV is defined as the sum of the hyperbolic excess speed at Earth with the absolute value of all of the $\Delta V_i$'s calculated at subsequent planetary encounters. Each $\Delta V_i$ is assumed to be impulsive, in the plane made by the arrival and departure velocity vectors at the ith encounter, and at the periapsis point with respect to each encounter. An exception is where an encounter is specified as an 'Intermediate Point' (IP). An IP is a point of zero mass with user-specified distance away from the centre of the ecliptic whose polar angles (θ, φ) are additional optimization parameters for the NOMAD solver. An IP corresponds to a Deep Space Maneuver (DSM). The $\Delta V_i$ at an IP is defined as the magnitude of change in velocity required at the IP to take the spacecraft from body i-1 to body i+1 via the IP – thus the notion of periapsis is not applicable for such an encounter. At destination, a flyby of 1I is assumed and so the final $\Delta V_i$ is zero. For deriving celestial body positions and velocities as a function of time, the NASA NAIF SPICE Toolkit is used and corresponding binary SPK files are utilised.

For all planetary encounters a minimum limit periapsis altitude of 200km is specified relative to the planet's equatorial radius. In the case of Jupiter, this equatorial radius is taken at the 1bar level, i.e. 71492km.

In this paper, we add results for the long-term evolution of the total ΔV for the trajectory Earth-Jupiter-solar Oberth maneuver-1I/'Oumuamua (E-J-3SR-1I), where SR denotes the solar radius. We analyse a proposed 2033 launch in greater detail, in particular in regard to finding the optimal Solar Oberth perihelion distance, which in fact shall be shown to be larger than 3SR.

For comparison purposes, it was decided that trajectories directly from Jupiter to 1I, *without* a Solar Oberth Maneuver between Jupiter and 1I, should be analysed, i.e. X–J-1I (where X is some combination of planetary encounters beginning at Earth). The reasoning for this is that as the heaviest planet in the Solar System, Jupiter is the planet best placed for achieving maximum benefit from the 'Oberth Effect'. The Oberth Effect holds that in the presence of a gravitational well, the total energy change of an orbiting spacecraft induced by a change in velocity ΔV of the spacecraft, reaches a maximum at the shortest distance and fastest speed with respect to the centre of attraction, i.e. at periapsis. For a spacecraft on an escape orbit, this change in energy manifests as a change in the hyperbolic excess speed of the body. Thus a high hyperbolic excess speed can be achieved for a low ΔV. For the high heliocentric speeds necessary to catch up with 1I/'Oumuamua, as large a gravitational well as possible must be chosen to make the most of the Oberth Effect.

Finally, by analysing missions from Mars to Jupiter to 1I/'Oumuamua (M-J-1I) (with Mars as the home planet) it was determined that a propitious relative positioning of Mars and Jupiter occurred with a rather late Jupiter arrival in 2032. This could potentially offer an opportunity for missions from Earth to Mars with a later launch than those detailed above and consequently allow a longer mission preparation time. Thus missions with Mars as the last visit before Jupiter and then 1I/'Oumuamua were investigated.

## 3. Results

### 3.1 Jupiter flyby – Solar Oberth maneuver at 3Solar Radii trajectories between 2020-2060

The original Project Lyra [26] paper discussed an Earth to Jupiter to 3 Solar Radii to 1I trajectory (E-J-3SR-1I), employing a Solar Oberth maneuver, launching in 2021. The favourable timing of this trajectory relies on:

(a) the propitious positioning of Jupiter (in its 11.9 year orbit around the Sun) relative to 1I.
(b) the close conjunction of Earth and Jupiter, at respectively launch and Jupiter arrival, such as to make optimal use of Earth's orbital velocity.

As is clear from Hein et al. [26], (a) and (b) conspire to allow a minimum total ΔV of around 18.3km/s with a launch in 2021.

Figure 1 shows the long-term results for the total ΔV for this E-J-3SR-1I trajectory. (In order to reduce the computation time required to produce this particular plot, accuracy in the calculation of ΔV was compromised.) As one might expect, there is a period of 12 or so years between successive coincidences of (a) and (b), with minima lasting for 4 years or so.

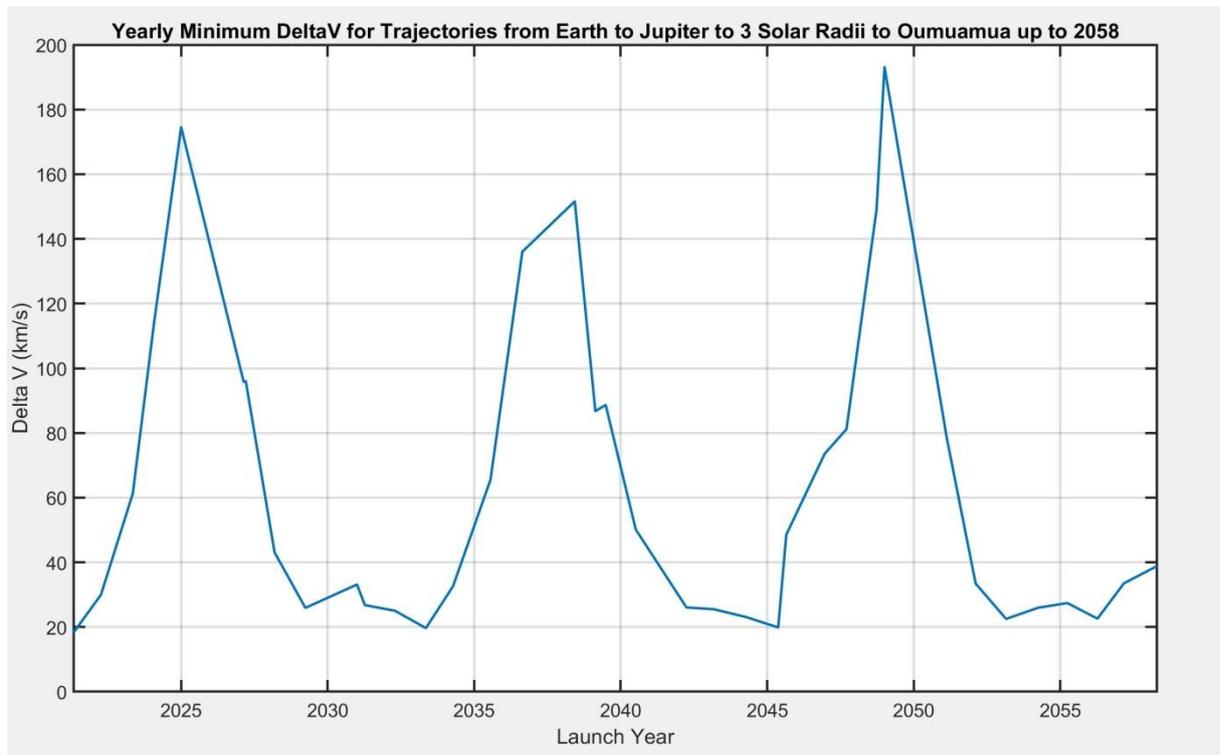

Figure 1: Minimum ΔV trajectories to 1I/'Oumuamua for E-J-3SR-1I

Note that the mission duration is not constrained for these calculations, so corresponding mission durations shown in Figure 2 are optimal ones in terms of ΔV. It can be seen that for a launch in 2033 and a Solar Oberth at 3 Solar radii, a flyby of 1I can be achieved with a mission duration of 16 years and a ΔV of roughly 20 km/s. Developing a spacecraft for a launch in 2033 is entirely within the range of typical development durations of interplanetary spacecraft and feasible from an engineering point of view.

As can be seen in Figure 1, even a much later launch date is possible in 2045 with a mission duration of 26 years. This would provide about 25 years for developing and launching a spacecraft. The corresponding ΔV for 3 Solar radii Solar Oberth is close to the 2033 launch of 20 km/s. A much later launch date in 2057 would be possible, however, it would exceed the maximum acceptable mission duration of 30 years by about 7 years. However, if longer mission durations are considered acceptable, a launch in 2057 would be another alternative.

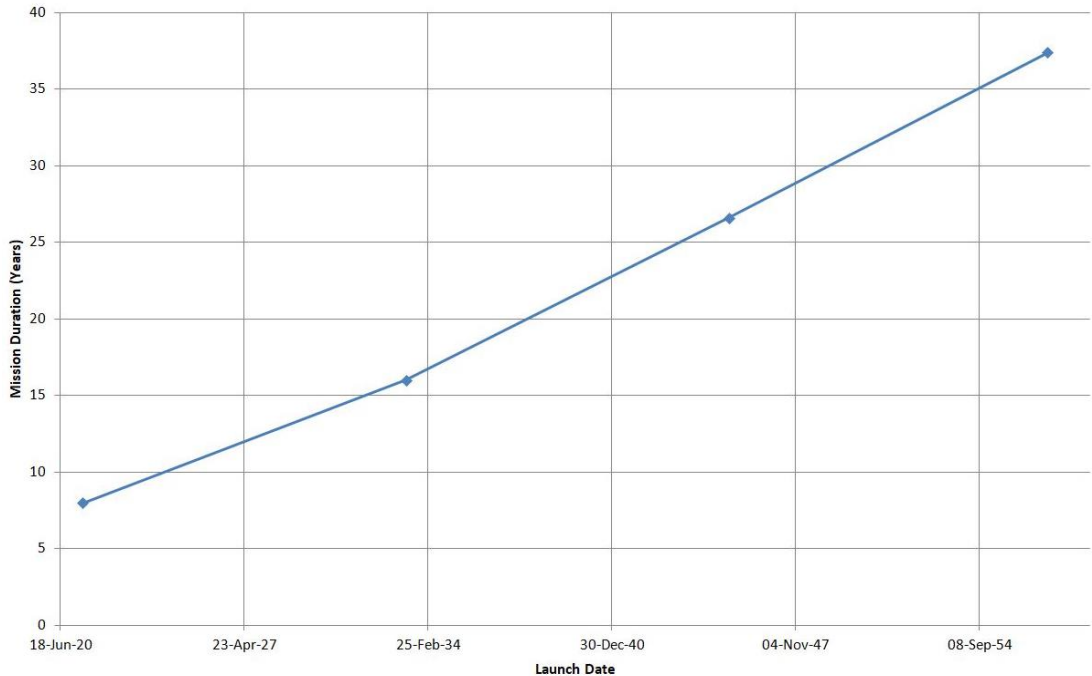

Figure 2: Mission duration versus launch date for minimum ΔV trajectories for four Jupiter cycles (E-J-3SR-1I)

To summarize, there are several additional opportunities of launching a spacecraft to 1I/'Oumuamua post 2024 with opportunities in 2033, 2045, and 2057. The ΔV of about 20 km/s could be achieved by a combination of powered Jupiter flyby and solar Oberth maneuver, using existing or near-term technologies, as has already been demonstrated in Hein et al. [26]. For example, a Falcon Heavy could launch a spacecraft of about 37 kg to 1I/'Oumuamua and a SLS launcher a 122 kg spacecraft, respectively.

## 3.2 Closer examination of 2033 mission and dependency on perihelion of Solar Oberth

The 3 solar radii perihelion distance for the Solar Oberth burn in [26], used in the previous Section's calculations was taken from the KISS architecture [29]. It seems reasonable to question whether this is an optimal value for the E-J-SO-1I trajectory, i.e. for the Oumuamua mission, which is under specific consideration here. Furthermore, the heat flux at 3 solar radii is about an order of magnitude higher compared to the one at the closest encounter of the Parker Solar Probe mission with the Sun. Hence, a larger distance from the Sun during the Solar Oberth burn would relax the requirement on the heat shield. As a consequence, 4 different E-J-SO-1I mission scenarios, (1), (2), (3) & (4) were explored. These are explained in Table 1.

Figure 3 demonstrates the relationship between total ΔV and perihelion distance measured in Solar Radii where 1 Solar Radius is taken as 695,700km for these 4 different mission scenarios.

| E-J-SO-1I mission scenario | Solar Oberth to Oumuamua Flight Time Upper Bound /days | Reasoning |
|---|---|---|
| (1) | 6000 | Approximately the optimal value observed in the 3 Solar Radii trajectory |
| (2) | 7500 | (1)+1500days |
| (3) | 10000 | (2)+2500days |
| (4) | 22000 | Allows intercept of Oumuamua to occur within the SPICE Kernel Data Set (i.e. < 2101AD) |

**Table 1: 4 Different mission scenarios for E-J-SO-1I**

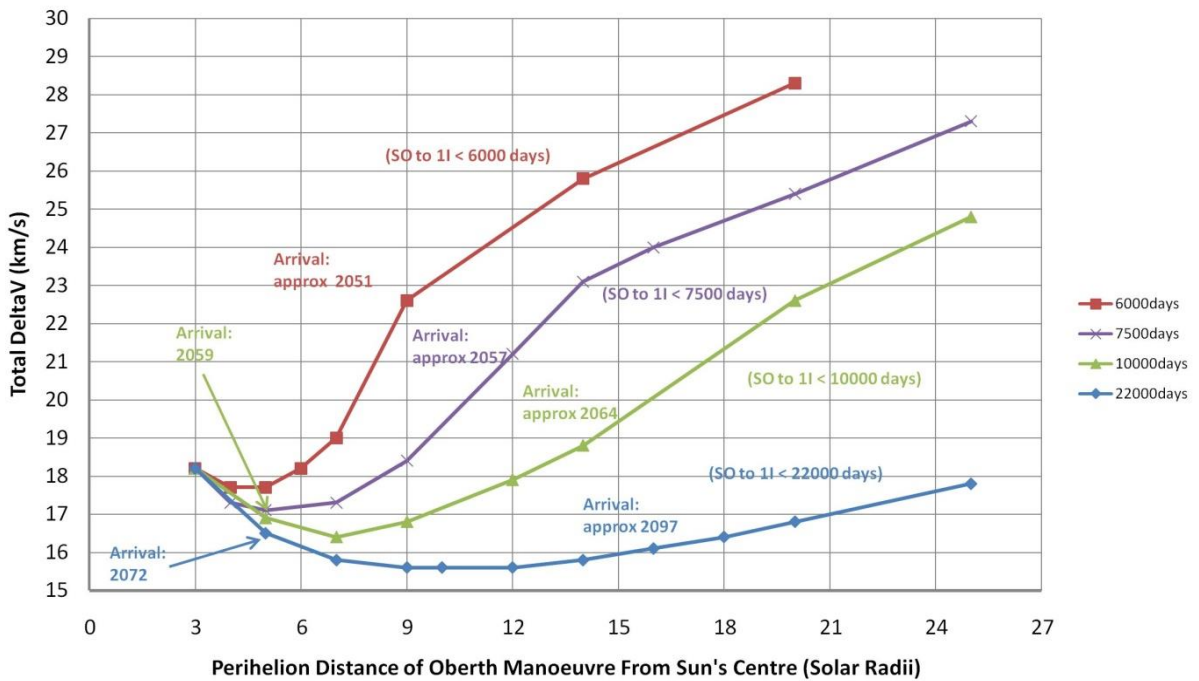

**Figure 3: Variation in total ΔV with increasing perihelion of Solar Oberth Maneuver for Scenarios 1 (red), 2 (violet), 3 (green) & 4 (blue)**

Examining Figure 3, there is initially a reduction in total ΔV as the perihelion of the Oberth starts to increase above 3 Solar radii for all scenarios. There is then a minimum (of around 4 Solar radii for (1), 5 Solar radii for (2), 7 Solar radii for (3) and 12 Solar radii for (4)) after which point the ΔV begins to climb. This initial reduction seems counter-intuitive as one would expect the effectiveness of the Solar Oberth to drop with increasing perihelion imposing a larger ΔV requirement on the Solar Oberth maneuver. The explanation for this phenomenon is clear when one takes into account the whole ΔV requirement, which importantly includes the ΔV at Jupiter perijove. The breakdown of the total ΔV into its three components, i.e. at Earth, at Jupiter perijove and at the Solar Oberth, is shown in Figure 4 for scenario (1) as an illustration.

Referring to Figure 4, the ΔV required at the Solar Oberth does indeed climb with increasing perihelion (the green portion), however there is a corresponding reduction in that required at perijove (the red portion) up to around 7 Solar radii and thereafter this contribution increases again. The combined effect is for an overall optimum to occur at around 4SR/5SR for this scenario (1).

Hereafter the scenario (1) at 6 Solar radii option shall be adopted (replacing the KISS value of 3 Solar radii) because, although the overall ΔV of 18.2km/s is 0.5km/s greater than the minimum at 4 Solar radii /5 Solar radii, nonetheless the ΔV requirement at Jupiter is zero thus possibly eliminating the necessity for an extra burn stage for the spacecraft, with presumably a corresponding reduction in mission complexity. There is also a useful reduction in solar heat flux by a factor of 2 or so, with implications in terms of lower heat shield mass.

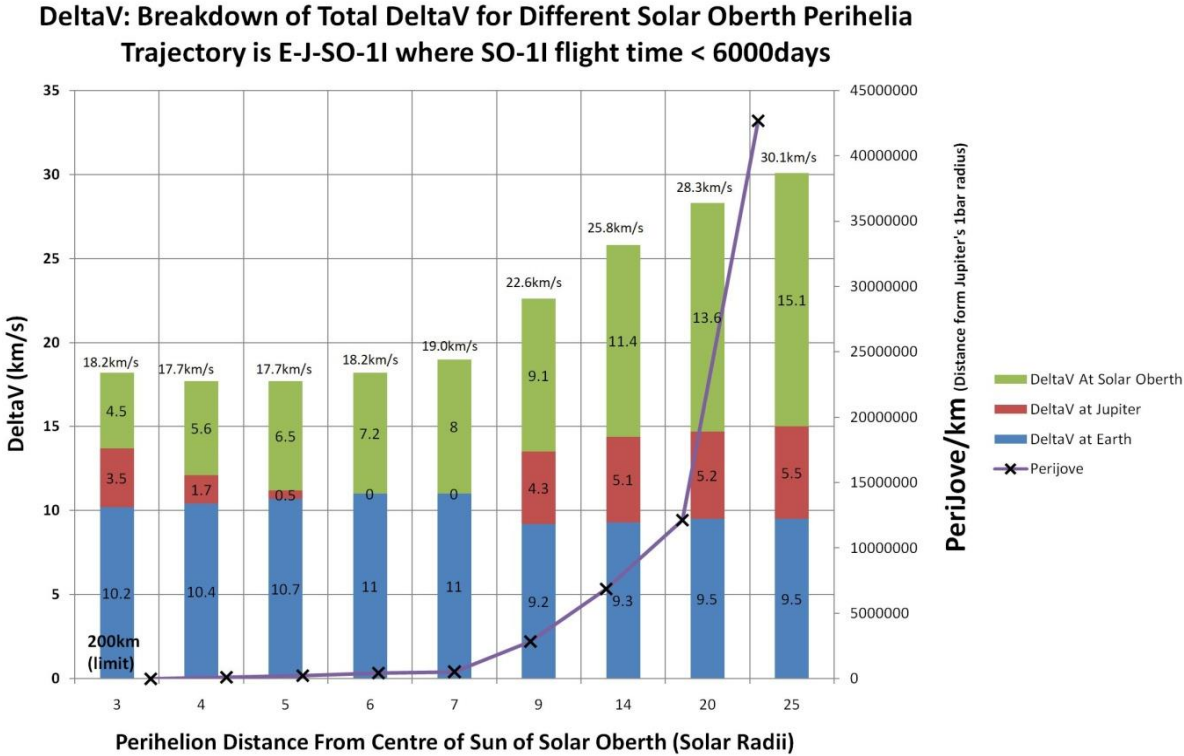

**Figure 4: Breakdown of total ΔV with increasing perihelion of Solar Oberth Maneuver for Scenario (1)**

## 3.3 Jupiter flyby – Solar Oberth Maneuver trajectories with Deep Space Maneuver

To use the architecture suggested by KISS [29] but with a Solar Oberth at 6 Solar radii, a DSM was inserted at 3.2 AU after Earth launch (launch being in 2030, i.e. 3 years before the optimal launch year found in the preceding section), then an Earth return followed by Jupiter, 6 Solar radii and then on to 1I. Thus E-DSM-E-J-6SR-1I. The E-DSM-E component represents a round trip of about 3years and is the reasoning behind setting the DSM at 3.2 AU. This yielded a total ΔV of 15.3km/s, an appreciable improvement on E-J-6SR-1I. The time from launch to 1I encounter would be about 22 years. Refer to Figure 5 for the E-DSM-E-J-6SR-1I trajectory.

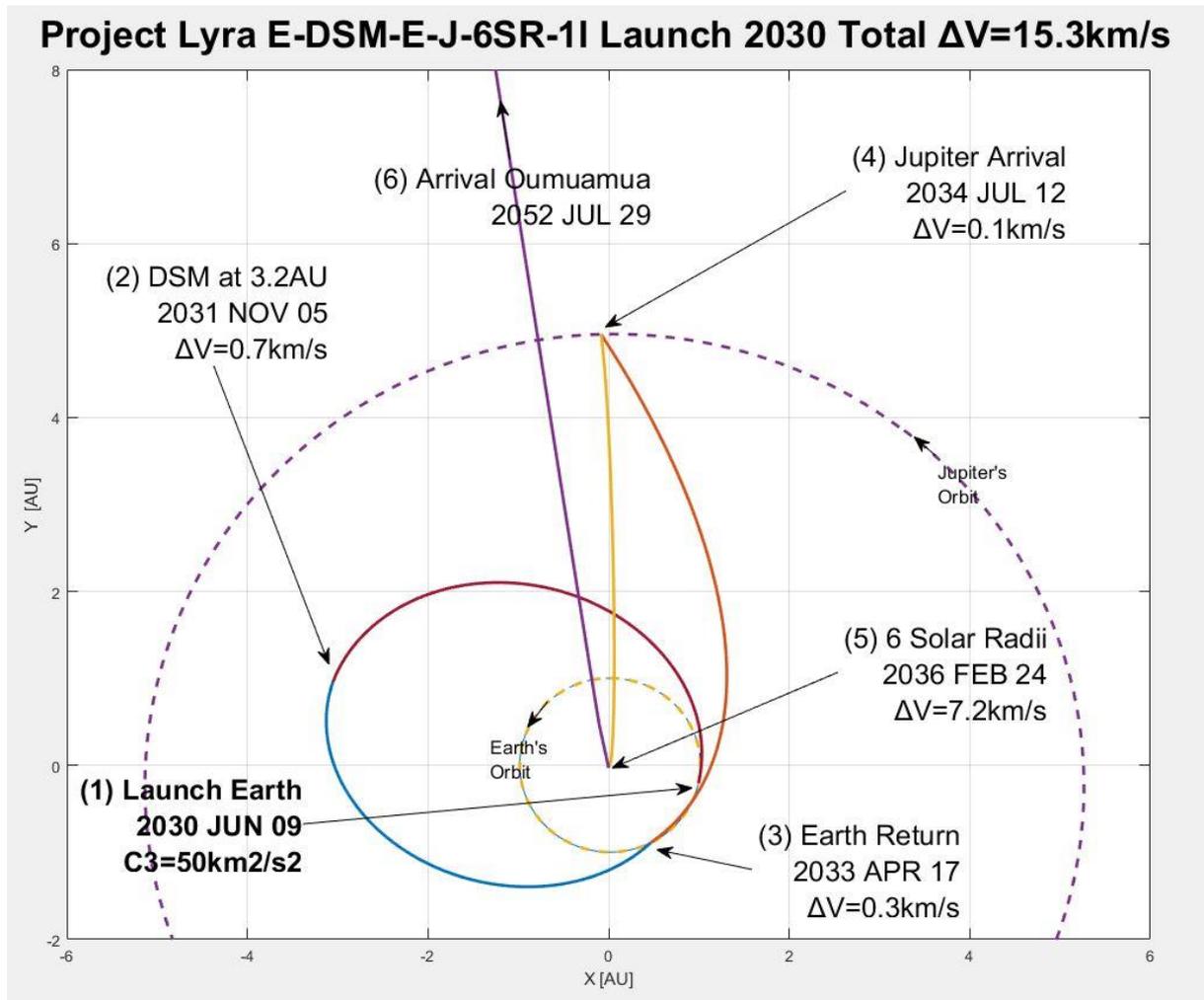

**Figure 5: E-DSM-E-J-6SR-1I trajectory with 2030 launch**

## 3.4 Direct Trajectories from Jupiter to 1I/'Oumuamua

The simplest mission sequence is Earth to Jupiter to 1I, i.e. E-J-1I with a Jupiter Oberth maneuver. A mission duration of 20 years was selected. Results indicate an optimal launch in 2031 February with a ΔV of around 26.5 km/s and arrival in 2051. We shall call this scenario (5):

Scenario (5): E-J-1I (Mission Duration=20years.)
Launch 2031 FEB 10
C3=96.77km2/s2
ΔV at Jupiter = 16.7km/s
Total ΔV = 26.5km/s

Ideally, this needs to be reduced to less than 18.2 km/s to make the mission a viable alternative to the E-J-6SR-1I scenario. Tackling this problem, it seems that the main hurdle is the large ΔV for the Jupiter Oberth maneuver, required to get to 1I. It appears that if this can be reduced and the majority of the ΔV can be placed at Earth, then this Earth ΔV can potentially be diminished by a combination of gravitational assists of the inner planets, i.e. Venus, Earth & Mars.

Can the plane of the orbit arriving at Jupiter be modified altering the Jupiter Oberth maneuver opening angle such as to enable lower ΔV values? The problem is that the trajectory from Earth to Jupiter is approximately in the ecliptic plane. An attempt to make the plane of the orbit from Earth to Jupiter outside of the ecliptic was explored. (This can be done in OITS by specifying the radial distance from the Sun of an 'Intermediate Point' [IP], otherwise known as a Deep Space Maneuver [DSM], between Earth and Jupiter, whose ecliptic polar angles are then optimized with the various other optimization parameters). This met with marginal success with a radial distance set to 4.9AU. Thus the E-J-1I orbit trajectory for launch 2031 had a ΔV of 26.5km/s whereas with a DSM at 4.9AU this reduced to around 24.3km/s:

Scenario (6): E-DSM-J-1I (Mission Duration=20years.)
Launch 2031 FEB 25
C3=94.26km2/s2
ΔV at DSM =2.4km/s
ΔV at Jupiter = 12.2km/s
Total ΔV = 24.3km/s

Clearly the ΔV at Jupiter has reduced but is there scope for reducing this further? To this end, a Jupiter return was considered - arrival at Jupiter from Earth is followed by a half Jupiter-year return of approximately 6years and achieving a distance of 5.2AU significantly out of the ecliptic. This Jupiter-Jupiter transfer could potentially have a large inclination and make inroads into the opening angle at the Jupiter return. Here are results for E-J-DSM-J-1I, Scenario (7).

Scenario (7): E-J-DSM-J-1I (Mission Duration 27years.)
Launch 2024 AUG 24
C3=141.6km/s
ΔV at Jupiter = 0km/s
ΔV at DSM =1.4km/s
ΔV at Jupiter Return = 9.4km/s
Total ΔV = 22.8km/s

This seems to have met with some success in that now the ΔV at Jupiter has significantly reduced and Earth C3 has increased. This looks promising in that a combination of inner planet gravitational assists which will improve on the ΔV to get to Jupiter can be sought.

Thus the sequence X-J-DSM-J-1I was investigated. Table 2 summarises these missions.

Here is such a trajectory with an Earth return after 1 year then heads off towards Jupiter and returns to Jupiter (E in Table 2):

Scenario (8): E-E-J-DSM-J-1I (Mission Duration 28 years.)
Launch 2023 AUG 22
C3=0km/s
ΔV at Earth Return= 5.6km/s
ΔV at Jupiter = 0.0km/s
ΔV at DSM =1.4km/s
ΔV at Jupiter Return = 9.4km/s
Total ΔV = 16.4km/s

Refer to Figure 6. Mission scenarios (7) and (8) both have return trips to Jupiter lasting 2200 days or so, i.e. half a Jupiter year. This allows a high inclination transfer orbit with respect to the ecliptic. However on further investigation of the X-J-DSM-J-1I combination, a benefit was also observed for return trips lasting a much shorter time than this - and with low inclinations with respect to Jupiter - refer to the trajectories (F) & (G) in Table 2. Note (G) is basically the same passage to Jupiter as the ESA planned JUICE mission in 2022 and is an extremely efficient trajectory – refer to Figure 7.

The results are obtained by using an additional error estimation code for OITS, as OITS assumes arrival at DSM between Jupiter arrival and return, is outside Jupiter's Sphere of Influence (SOI) and travelling with a small velocity relative to Jupiter. As Jupiter's SOI is extremely large, i.e. Laplace SOI=0.3AU from Jupiter, and in fact on returning to Jupiter, the spacecraft has been loitering within this SOI and has a significantly non-zero component. The ΔV's calculated by OITS have to be increased by around 4 km/s at the DSM for these '-J-DSM-J-1I' trajectories. This error has been taken into account for (F) and (G).

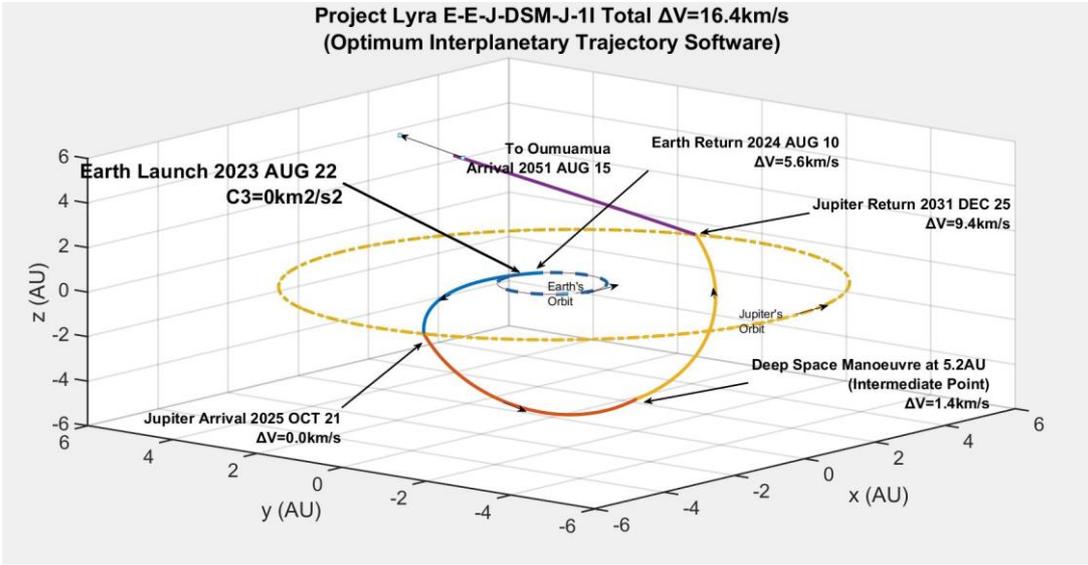

**Figure 6: E-E-J-DSM-J-1I trajectory**

Note that although scenario (8), i.e. that shown in Figure 6 and trajectory (E) in Table 2, has a low ΔV of 16.4km/s, one key drawback is the launch date in 2023, which is too short notice for developing and launching a spacecraft, using current engineering practice. Furthermore, although the time from launch to 1I/'Oumuamua encounter of 28 years is within the limits of 30 years, it is clearly less attractive than the durations from Section 3.2.

### 3.5 Missions to Mars followed by Jupiter and then 1I/'Oumuamua

A fruitful sequence appears to include Mars flybies, i.e. E-M-DSM-M-J-1I., with a DSM at approximately Mars' radial distance, i.e. 1.523AU. Thus after arriving at Mars from Earth, some time is spent loitering in the vicinity of Mars before a return and then an encounter with Jupiter. Refer to trajectories (H) & (I) in Table 2.

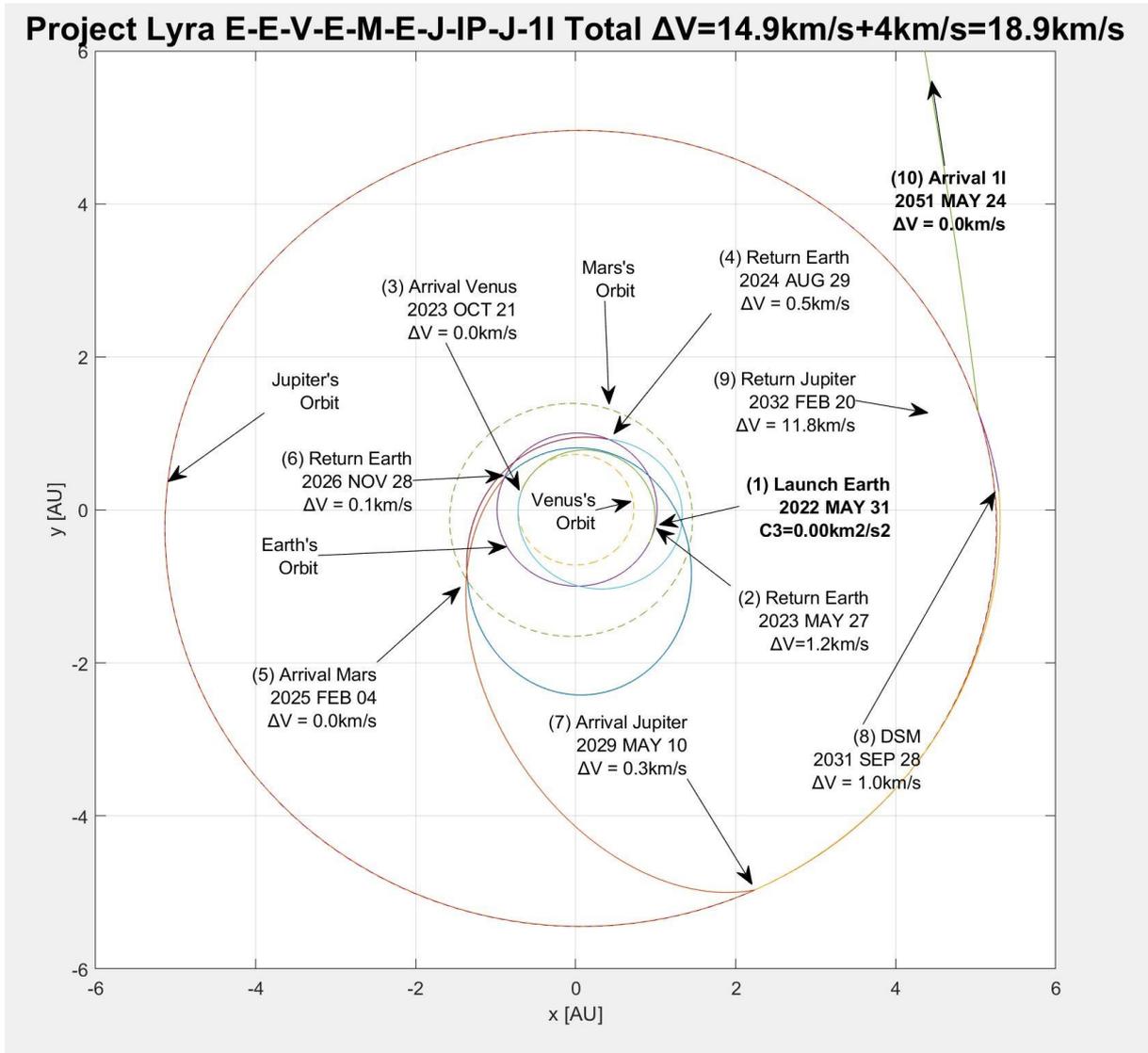

**Figure 7: E-E-V-E-M-E-J-IP-J-1I trajectory**

## 3.6 Summary of Results

Table 2 provides an overview of the trajectories presented in this paper. The launch dates are between 2022 and 2033 and the arrival dates at 1I/'Oumuamua are between 2051 and 2052.

**Table 2: Summary of scenarios**

| Trajectory (Scenario) | Mission | Launch Earth | Arrival at 1I | Arrival At Jupiter prior to 1I | Loiter Time [days] | Mission Duration [years] | C3 at Earth [km²/s²] | Total ΔV [km/s] | ΔV Prior to First Jupiter Encounter [km/s] | ΔV After and including first Jupiter Encounter [km/s] |
|---|---|---|---|---|---|---|---|---|---|---|
| A | E-J-6SR-1I | 2033 MAY 08 | 2052 DEC 09 | 2034 JUL 30 | N/A | 19 | 121 | 18.2 | 11.0 | 7.2 |
| B (1) | E-DSM[a]-E-J-6SR-1I | 2030 JUN 09 | 2052 JUL 29 | 2034 JUL 12 | N/A | 22 | 50 | 15.3 | 8.0 | 7.3 |
| C (5) | E-J-1I | 2031 FEB 10 | 2051 FEB 05 | 2033 AUG 09 | N/A | 20 | 97 | 26.5 | 9.8 | 16.7 |
| D (7) | E-J-DSM-J-1I | 2024 AUG 24 | 2051 AUG 18 | 2031 DEC 22 | N/A | 27 | 142 | 22.7 | 11.9 | 10.8 |
| E[c] (8) | E-E-J-DSM-J-1I | 2023 AUG 22 | 2051 AUG 15 | 2031 DEC 25 | N/A | 28 | 0.0 | 16.4 | 5.6 | 10.7 |
| F[c] | E-E-DSM[b]-E-J-DSM-J-1I | 2024 OCT 01 | 2051 SEP 25 | 2032 MAR 22 | 850[d] | 27 | 0.0 | 20.7 | 3.7 | 16.9 |
| G[c] | E-E-V-E-M-E-J-DSM-J-1I | 2022 MAY 31 | 2051 MAY 24 | 2029 MAY 10 | 1017[d] | 29 | 0.0 | 18.9 | 2.2 | 16.7 |
| H | E-M-DSM-M-J-1I | 2028 DEC 15 | 2052 DEC 09 | 2032 MAR 05 | 417[e] | 24 | 9.4 | 21.9 | 10 | 11.9 |
| I | E-E-M-DSM-M-J-1I | 2027 DEC 13 | 2051 DEC 07 | 2032 FEB 08 | 471[e] | 24 | 0.0 | 19.8 | 11 | 8.8 |

[a] Deep Space Maneuver is at 3.2AU from Sun.

[b] First Deep Space Maneuver is at 2.2AU from Sun.

[c] Earth return 1 year after launch i.e. E-E-

[d] Loiter time in vicinity of Jupiter

[e] Loiter time in vicinity of Mars

From Table 2, it can be seen that there is a tradeoff between total ΔV and mission duration. This is demonstrated in Figure 8, where scenario E-J-6SR-1I (trajectory (A) & launch in 2033) and E-DSM-E-J-6SR-1I (trajectory (B) & launch in 2030) toward the bottom left of the figure seem good candidates for a mission to 'Oumuamua. Both scenarios are on the Pareto frontier, indicated as a blue line. Both scenarios are on the Pareto frontier, indicated by the solid brown line and a dashed blue line in Figure 8 for different minimum distances from the Sun. They would also allow for ample time for spacecraft development. The scenarios (1) 4 Solar radii / 5 Solar radii, 5 Solar radii (Arrival 2048), and (1) 3 Solar radii are not considered here due to their proximity to the Sun but this decision would depend on the specific considerations of designers of the mission, should it arise.

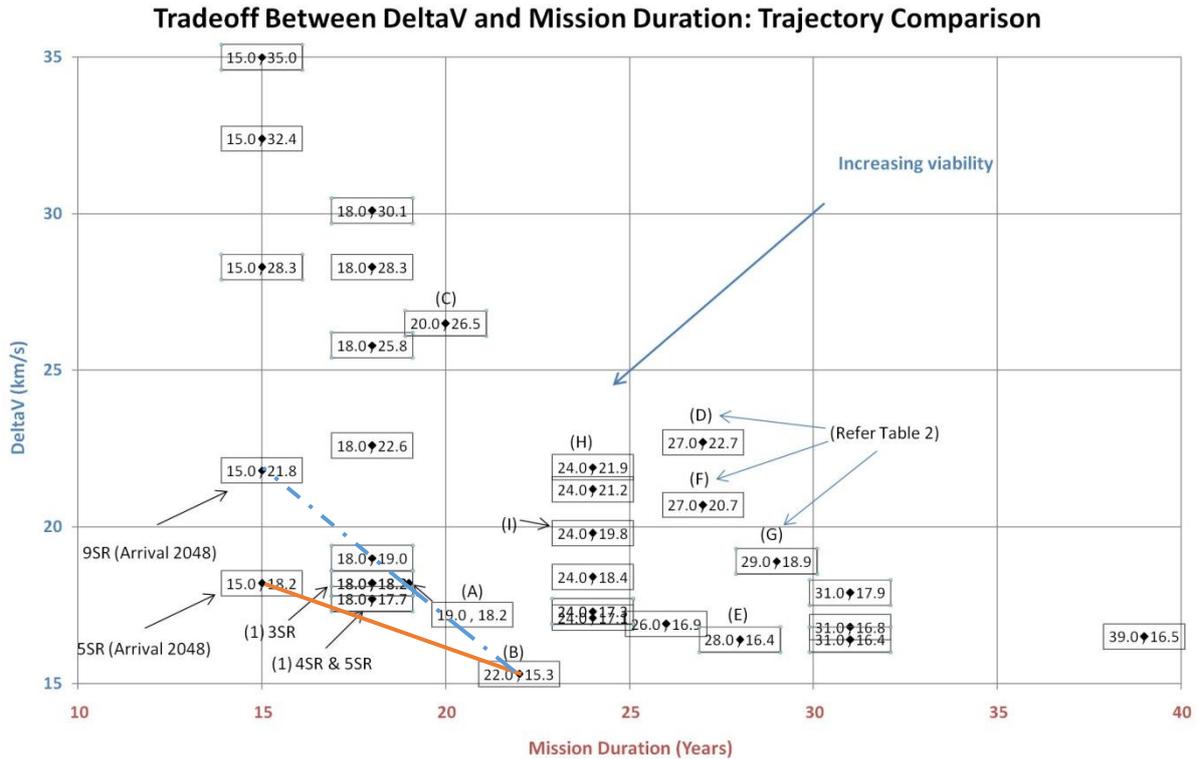

**Figure 8: ΔV versus mission duration for encounters with 1I between 2048 and 2052 The dashed blue line is the Pareto frontier with a minimum Oberth distance of 6 Solar radii and the solid brown Pareto frontier line extends this limit to 5 Solar radii.**

## 4. Conclusions

This paper presented a feasibility analysis of a mission to 1I/'Oumuamua using current and near-term technologies with a launch date in the mid-2020s to mid-2030s, which would allow for sufficient time for spacecraft development. We identify at least two missions to 1I/'Oumuamua with launch dates between 2030 and 2033 with total velocity increments of 15.3 km/s and 18.2 km/s, and an arrival at 1I/'Oumuamua in 2048. These launch dates also provide about a decade for spacecraft development from 2019 onwards, in contrast to the previously identified launch dates in the 2020-2021 timeframe, where only 1-2 years are left for spacecraft development. Furthermore, we demonstrate that these missions can be performed with the closest distance to the Sun for the Oberth burn at 5 solar radii, resulting in heat flux values which are of the same order as for the Parker Solar Probe. For future work, we propose a conceptual design of the associated spacecraft and corresponding instrument suite.